\documentclass[conference]{IEEEtran}

\usepackage{latexsym,bm}
\usepackage[tbtags]{amsmath}

\usepackage{algorithm}
\usepackage{algpseudocode}

\usepackage{amssymb}
\usepackage{cite}
\usepackage{stfloats}

\usepackage{multirow}
\usepackage{booktabs}
\usepackage{amsthm}
\usepackage{float}
\usepackage{amsthm}
\usepackage{amssymb}

\usepackage{graphicx}
\usepackage{subfigure}
\usepackage[keeplastbox]{flushend}

\begin{document}
\title{A Two-Phase Power Allocation Scheme for CRNs Employing NOMA}
%\\ for IEEE Conferences}

\author{\IEEEauthorblockN{Ming Zeng,
Georgios I. Tsiropoulos\IEEEauthorrefmark{1},
Animesh Yadav,
Octavia A. Dobre, and
Mohamed H. Ahmed}
\IEEEauthorblockA{Faculty of Engineering and Applied Science, Memorial University, St. John, Canada}
\IEEEauthorblockA{\IEEEauthorrefmark{1}School of Electrical and Computer Engineering, National Technical University of Athens, Greece}
%\IEEEauthorblockA{\IEEEauthorrefmark{3}Faculty of Engineering and Applied Science, Memorial University, St. John, Canada}

\IEEEauthorblockA {Email: \{mzeng, animeshy, odobre, mhahmed\}@mun.ca, gitsirop@mail.ntua.gr}}


%Telephone: (800) 555--1212, Fax: (888) 555--1212}


\maketitle

\begin{abstract}
In this paper, we consider the power allocation (PA) problem in cognitive radio networks (CRNs) employing non-orthogonal multiple access (NOMA) technique. Specifically, we aim to maximize the number of admitted secondary users (SUs) and their throughput, without violating the interference tolerance threshold of the primary users (PUs). This problem is divided into a two-phase PA process: a) maximizing the number of admitted SUs; b) maximizing the minimum throughput among the admitted SUs. To address the first phase, we apply a sequential and iterative PA algorithm, which fully exploits the characteristics of the NOMA-based system. Following this, the second phase is shown to be quasiconvex and is optimally solved via the bisection method. Furthermore, we prove the existence of a unique solution for the second phase and propose another PA algorithm, which is also optimal and significantly reduces the complexity in contrast with the bisection method. Simulation results verify the effectiveness of the proposed two-phase PA scheme.
\end{abstract}


% \fi
\IEEEpeerreviewmaketitle

\section{Introduction}
With the proliferation of smart mobile devices, such as smart phones, M2M, and emerging wearables, global mobile data traffic is expected to grow to 30.6 EB per month by 2020 \cite{a1}. In order to meet the mobile data traffic requirement, a 10 Gbps peak data rate and 1 Gbps user experienced data rate are proposed to be supported by 5G \cite{a2}. Hence, with limited spectrum availability, enhancing spectral efficiency is of significant importance for 5G, becoming one of its main design requirements.

A prevailing way to address spectrum scarcity is to apply dynamic and efficient spectrum accessing techniques, such as cognitive radio (CR) \cite{a3,a4,a5, 25}. CR networks (CRNs) are envisioned to provide more bandwidth to mobile users through heterogeneous architectures and dynamic spectrum access techniques. Therefore, network users are divided into two main groups: licensed/primary users (PUs) and unlicensed/secondary users (SUs). Correspondingly, two requirements have to be satisfied in CRNs: a) the interference introduced by the operation of SUs towards PUs should be kept under a certain threshold, and b) the admitted SUs should meet their minimum data rate requirement. Particularly, for spectrum sharing, spectrum underlay or overlay techniques can be used for designing CRNs \cite{a3,a4}.

Another way is to employ non-orthogonal multiple access (NOMA) \cite{22, 23, 24, a19}, which has attracted considerable attention recently owing to its potential to achieve superior spectral efficiency. Unlike conventional orthogonal multiple access (OMA), NOMA multiplexes users in the power-domain at the transmitter side, and conducts multi-user signal separation using successive interference cancellation (SIC) at the receiver side. Thus, in NOMA-based systems, power allocation (PA) is of great importance, since it not only impacts the users' achievable data rates, but also determines their channel access. A variety of PA strategies have been proposed so far, targeting different aspects of PA in NOMA \cite{a10, a12, a16, a17}. CR-inspired PA is adopted in \cite{a16}, where NOMA is considered as a special case of CRNs and the user with poor channel condition (poor user) is viewed as a PU. This way, the quality of service (QoS) for the poor user can be strictly guaranteed. However, the performance of the user with better channel condition may be sacrificed since this user is served only after the poor user's QoS is met. To offer more flexibility in the tradeoffs between the user fairness and system throughput, a dynamic PA scheme is proposed in \cite{a17}, which strictly guarantees the performance gain of NOMA over OMA for both poor user and user with better channel conditions.

%The criterion of maximizing the multi-user proportional fairness (PF) is adopted in \cite{a12}, and the PA problem is formulated as a weighted sum of user throughput. Although optimal performance can be obtained, the PA is computationally complex and feedback of the determined allocated power to the respective users is required. Thus, two simple, suboptimal PA methods were proposed in \cite{a12}, i.e., a fixed (channel-independent) PA and a fractional transmit power control (FTPC). While the computational complexity of FTPC is as low as the fixed (channel-independent) PA, it achieves a better performance, since it allocates power proportional to the path-loss. However, the performance of FTPC is still not good enough due to the fact that allocating the transmission power proportional to the path-loss is not flexible, and it remains a problem how to select the appropriate proportion ratio.

In this paper, we study the two concepts of NOMA and CRN, i.e., a CRN employing NOMA for its SUs, leading to a further increase in spectral efficiency. In such a NOMA-based CRN, PA for SUs not only determines the channel access of SUs, but also affects the performance of PUs. Consequently, the performance of the adopted PA scheme is vital, and a full exploitation of the power domain should be achieved. Motivated by this, our contributions can be summarized as follows:
\begin{enumerate}
  \item we propose a two-phase PA scheme to maximize the number of admitted SUs and their throughput; 
  \item the first phase maximizes the number of admitted SUs. NOMA has been well studied in terms of network throughput, link quality, outage probability estimation etc. However, the system capacity in terms of number of admitted users has not been studied so far. The analysis in this paper provides useful insights;
  \item the second phase aims to maximize the minimum data rate among the admitted SUs. It is worth noting that the max-min problems are investigated in \cite{a21, a9}. \cite{a21} considers the max-min fairness criterion under statistical channel state information (CSI), and aims to achive outage balancing among the users, which is different from the user admission problem considered in this paper. In \cite{a9}, there is no QoS requirement for the users, and thus, sum rate maximization is pursued, which differs from our system model, in which each SU has a minimum data rate requirement;
  \item numerical simulations are conducted to verify the effectiveness of the proposed two-phase PA scheme.
\end{enumerate}

The rest of the paper is organized as follows. The system model and problem formulation are presented in Section II. The proposed two-phase PA process is introduced in Section III. Performance evaluation results are illustrated in Section IV and conclusions are drawn in Section V.

\section{System Model and Problem Formulation}

\subsection{System Model}
A hierarchical spectrum sharing CRN is considered, where spectrum underlay is employed. NOMA is adopted to reduce the interference among SUs so as to further improve the spectrum efficiency.

Fig. 1 shows the structure of the considered NOMA-based CRN. The PUs are served by the base station (BS) in the downlink. Meanwhile, the BS can serve the SUs simultaneously obeying the access rules (i.e., the interference from the SUs towards each PU should be less than a certain interference threshold). On the other hand, the admitted SUs should meet their minimum data rate requirement, which is characterized by the signal-to-interference-plus-noise ratio (SINR); an SU is admitted if its SINR requirement is met.

The system model and transmission settings follow the ones in \cite{a14}, and are defined as follows. We assume that there exist $M$ PUs and $N$ SUs in the network. The channel gain between the BS and the $n$th SU is denoted as $G_{n}, n \in \{1,\cdots,N \}$, which strongly depends on the distance between them. Likewise, we denote the channel gain from the BS to the $m$th PU as $g_{m}, m\in \{1,\cdots,M \}$. We consider that the channel gains are known at the BS. Without loss of generality, we arrange the SUs in a descending order as follows:
\begin{equation} \label{eq:eps1}
G_1 \geq \ldots \geq G_n \ldots \geq G_N.
\end{equation}

The aggregate noise and interference from all PUs towards the $n$th SU are denoted as $N_{n}$. As NOMA is employed among SUs, SUs with better channel gains can cancel the interference from users with lower channel gains through SIC. As a result, the SINR of SUs can be calculated as \cite{a18, a19, a20}
\begin{equation}\label{eq:eps3}
    \gamma_n=\frac{P_n G_n}{\sum_{j=1}^{n-1} {P_j G_n}+N_n},
\end{equation}
where $\gamma_n$ and $P_n$ denote the SINR and allocated power of the $n$th SU, respectively.

\begin{figure}
\centering
\includegraphics[width=0.5\textwidth]{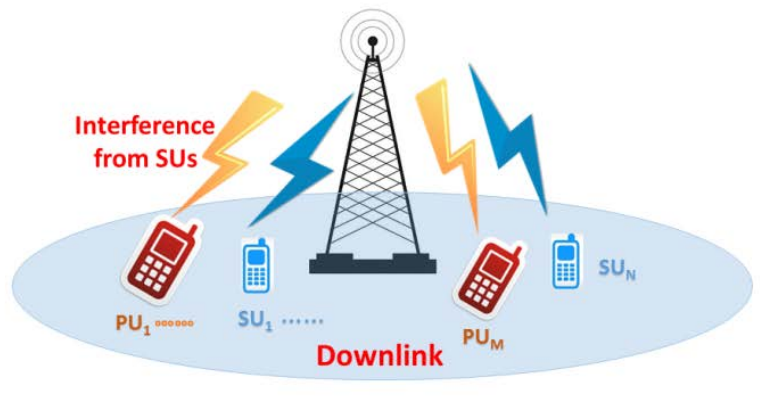}
\caption{Spectrum sharing scheme.}
\end{figure}

Further, if the SINR threshold of the $n$th SU is $\Gamma_n$, $\gamma_n$ needs to satisfy
\begin{equation}\label{eq:eps4}
    \gamma_n\geq \Gamma_n.
\end{equation}

In terms of the QoS requirement for PUs, denote the maximum interference level tolerable by the $m$th PU by $I_m$, the corresponding interference constraint can be formulated as
\begin{equation}\label{eq:eps5}
    \sum_{n=1}^{N}{P_n}g_m \leq I_m.
\end{equation}

Equation (\ref{eq:eps5}) can be further rewritten as
\begin{equation}\label{eq:eps6}
    P_s \leq \frac{I_m}{g_m},
\end{equation}
where $P_s=\sum_{n=1}^{N}{P_n}$ corresponds to the overall power for all SUs. This is fulfilled if
\begin{equation}\label{eq:eps9}
P_s \leq P_M,
\end{equation}
where $P_M=\min{ \big(\frac{I_m}{g_m}\big),m \in \{1,\ldots,M \}}$. As $P_s$ should also be constrained to a maximum power $P_{max}$, i.e., $P_s\leq P_{max}$, it yields
\begin{equation}\label{eq:eps10}
P_s = \min {(P_M,P_{max})}.
\end{equation}

In the following sections, $P_s$ is assumed to be known and directly used as the overall power constraint for SUs.

\subsection{Problem Formulation}
%The problem of maximizing the number of admitted SUs and their throughput is investigated. Assume that the maximum number of SUs which can be admitted is $L, L\in \{0,\ldots, N \}$. Further, the $L$ admitted SUs are denoted as $a_1, a_2, \dots, a_{L}$.\footnote{In the formulated problem, $L$ and the admitted SUs are parameters to be obtained, and are not known in advance.} On this basis, the objective function can be formulated as
%\begin{IEEEeqnarray*}{clr}\label{eq:user}
%\displaystyle\underset{\vec{P}}{\text{max}}   & \quad \min (\vec{\gamma}) \IEEEyesnumber \IEEEyessubnumber* \\
%\text{s.t.}
%& \quad \gamma_n \geq \Gamma_n,  n\in \{ a_1, a_2, \dots, a_{L} \}, \\
%   & \quad \sum_{n=a_1}^{a_{L}}{P_n} \leq P_s,
%\end{IEEEeqnarray*} 

%where $\vec{P}=[P_{1} \ldots P_{N}]$ and $\vec{\gamma}=[\gamma_{a_1} \ldots \gamma_{a_L}]$ are the overall PA vector and the SINR of the $L$ admitted SUs, respectively. The formulated objective function aims to maximize the minimum SINR among the $L$ admitted SUs. It is straightforward to decompose this problem into two phases: a) maximizing the number of admitted SUs under the QoS requirements for both the SUs and PUs; b) maximizing the minimum SINR among the admitted SUs via the allocation of the remaining power.

The problem of maximizing the number of admitted SUs and their throughput is investigated. This problem consists of two phases: 1) maximizing the number of admitted SUs under the PUs' QoS requirement and SUs' minimum data requirement; and 2) maximizing the minimum SINR among the admitted SUs via the allocation of the remaining power. Denote the total number of admitted SUs as $L, L\in \{0,\ldots, N \}$, and the corresponding SUs as $a_1, a_2, \dots, a_{L}$. On this basis, the problem in the first phase is formulated as
\begin{IEEEeqnarray*}{clr}\label{eq:user1}
\displaystyle\underset{\vec{P}}{\text{max}}  & \quad L \IEEEyesnumber \IEEEyessubnumber* \\
\text{s.t.}
& \quad \gamma_n \geq \Gamma_n,  n\in \{ a_1, a_2, \dots, a_{L} \}, \\
   & \quad \sum_{n=a_1}^{a_{L}}{P_n} \leq P_s.
\end{IEEEeqnarray*} 
where $\vec{P}=[P_{1} \ldots P_{N}]$ is the overall PA vector.
%\begin{subequations}
%\begin{align}
%  &{}^{\displaystyle{\max \: L } }_{\vec P} \\
%  \text{subject to}\nonumber \\
%     & \gamma_n\geq \Gamma_n,  n\in \{ a_1, a_2, \dots, a_{L} \} \\
%   & \sum_{i=a_1}^{a_L}{P_n} \leq P_s.
%\end{align}
%\end{subequations}

After the PA process of the first phase, SUs admission is ascertained. We use $L^ \star$ and $a_1, a_2, \dots, a_{L^ \star}$ to denote the maximum number of admitted SUs and the corresponding SUs. On this basis, the problem in the second phase is to further increase the SINR among the $L^ \star$ admitted SUs to increase system throughput, which is formulated as
%\begin{subequations}
%\begin{align}
%&{}^{\displaystyle{\max  \: \min (\vec{\gamma}) } }_{\vec P}  \\
%\text{subject to}\nonumber \\
%    &\gamma_n \geq \Gamma_n,  n\in \{ a_1, a_2, \dots, a_{L} \} \\
%   &\sum_{i=a_1}^{a_L}{P_n} \leq P_s
%\end{align}
%\end{subequations}
\begin{IEEEeqnarray*}{clr}\label{eq:user2}
\displaystyle\underset{\vec{P}_{L^ \star}}{\text{max}}  & \quad \min (\vec{\gamma}) \IEEEyesnumber \IEEEyessubnumber* \\
\text{s.t.}
& \quad \gamma_n \geq \Gamma_n,  n\in \{ a_1, a_2, \dots, a_{L^ \star} \}, \\
   & \quad \sum_{n=a_1}^{a_{L^ \star}}{P_n} \leq P_s,
\end{IEEEeqnarray*} 
where $\vec{P}_{L^ \star}=[P_{a_1} \ldots P_{a_{L^ \star}}]$ and $\vec{\gamma}=[\gamma_{a_1} \ldots \gamma_{a_{L^ \star}}]$ are the PA vector and the SINR of the ${L^ \star}$ admitted SUs, respectively.

We can consider the second phase from two perspectives: a) from the angle of exploiting the remaining power after the PA process of the first phase, with the goal of allocating the remaining power appropriately to maximize the minimum SINR among the ${L^ \star}$ admitted SUs; b) consider itself as an independent problem without taking into account the PA process of the first phase and its remaining power. In this case, (9) can be viewed as a problem of allocating the overall power of all SUs among the ${L^ \star}$ admitted SUs, subjecting to each admitted SU satisfying its SINR requirement.

\section{Proposed Two-Phase PA Scheme}
In this section, the proposed two-phase PA scheme is presented. First, we address the first phase and give a detailed description of the corresponding PA algorithm. Following this, the second phase is resolved.

\subsection{SUs Admission and Initial PA Algorithm}
In order to maximize the number of admitted SUs, we employ the sequential and iterative PA scheme, which makes full use of the characteristics of NOMA-based CRN and allocates power to SUs in a descending order according to their channel gains. 

According to the SINR requirements. i.e., (2) and (3), we have the following equations
%\begin{subequations}
\begin{align}
   P_n &\geqslant \Gamma_n \sum_{j=1}^{n-1}{P_j} +\frac{\Gamma_n N_n}{G_n}, n=1,\dots,N,  
\end{align}
%\end{subequations}
where the only variable is $\sum_{j=1}^{n-1}{P_j}$, since other parameters, i.e., $\Gamma_n$, $N_n$ and $G_n$ are known to the BS. Therefore, if we assign the power among SUs following the ascending order, i.e., from the $1$st SU to the $N$th SU sequentially, the power for the $n$th SU can be obtained easily, as $\sum_{j=1}^{n-1}{P_j}$ is already known. Specifically, the power for the $1$st SU is calculated as

\begin{equation}\label{eq:eps13}
  P_1=\frac{\gamma_1 N_1}{G_1}.
\end{equation}

Sequentially and iteratively, since the power of the $1$st SU is known, it is used for the PA of the $2$nd SU. We attain the following equation according to (10)
\begin{equation}\label{eq:eps14}
 P_2=\Gamma_2P_1+\frac{\Gamma_2 N_2}{G_2}.
\end{equation}

Similarly, the power for the $n$th SU is given by
\begin{equation}\label{eq:eps15}
   P_n=\Gamma_n \sum_{j=1}^{n-1}{P_j} +\frac{\Gamma_n N_n}{G_n}.
\end{equation}

Obviously, (13) can be used for the PA of all SUs. On the other hand, note that $P_s$ has not been considered during the above PA process. Therefore, during the PA for the $n$th SU, we also need to guarantee that the total power assigned to SUs, $\sum_{j=1}^{n}{P_j}$, does not exceed $P_s$. This is achieved by
\begin{equation}\label{eq:eps15}
    P_n=\min{(\Gamma_n \sum_{j=1}^{n-1}{P_j}+\Gamma_n \frac{N_n} {G_n},\: P_s-\sum_{j=1}^{n-1}{P_j})}.
\end{equation}

Furthermore, during each allocation, whenever $P_s-\sum_{j=1}^{n-1}{P_j}<\Gamma_n \sum_{j=1}^{n-1}{P_j}+\Gamma_n \frac{N_n} {G_n}$, it indicates there is not enough power left for the $n$th SU to meet its SINR requirement. Consequently, the PA process terminates and the $n$th SU to the $N$th SU receives no power to ensure that the QoS requirements for PUs are not violated. The admitted SUs are $1$st SU, $\dots$, $(n-1)$th SU, with the allocated power given by (13). After the PA process, note that there exists some remaining power, i.e., $P_s-\sum_{j=1}^{n-1}{P_j}$. While this power is not large enough to admit an extra SU, it can be further allocated to the admitted SUs to increase their SINR, and thus, to enhance the throughput. Particularly, according to (13), the power required for the admission of the $n$th SU is even larger than the sum of the power for all the former $n-1$ SUs, in case of $\Gamma_n \geq 1$. Therefore, the remaining power indeed randomly lies in the boundary $[0, P_n]$, which may help enhance the throughput of the former $n-1$ SUs significantly. 

In \cite{a14}, we have proven that when the SINR requirement of each SU is the same, the above PA algorithm is optimal. Moreover, the computational complexity is only $O(N)$.

\subsection{Maximize the Minimum SINR among the Admitted SUs}
\subsubsection{Analysis via Convex Optimization}
After the initial PA process, the SUs admission is done. On this basis, the second phase aims to maximize the minimum SINR among the admitted SUs.

Indeed, the second phase is quasiconvex. Note that the variables in (9) are the power values of the admitted SUs. Accordingly, the SINR of each SU is in the form of linear fractional function, which is a quasiconcave function. Following this, the operation of selecting the minimum value from a set of quasiconcave function will not change its quasi-concavity. Moreover, maximizing a quasiconcave function is equivalent to minimizing a quasiconvex function. In addition, the constraint functions are all convex. Consequently, the objective function is a quasiconvex function. According to \cite{a15}, a general approach to quasiconvex optimization relies on the representation of the sublevel sets of a quasiconvex function via a family of convex inequalities. Now by introducing an axillary variable $t$, (9) can be equivalently represented as
\begin{IEEEeqnarray*}{clr}\label{eq:P}
\displaystyle {\text{find}}  & \quad \vec{P}_{L^ \star} \IEEEyesnumber \IEEEyessubnumber* \\
\text{s.t.}
&\quad {\gamma_n} \geq t, \: \: n\in \{ 1,\ldots, {L^ \star} \} \\
   &\quad \gamma_n \geq \Gamma_n,n \in \{1,\ldots,{L^ \star} \} \\
    &\quad \sum_{n=1}^{{L^ \star}}{P_n}\leq P_s.
\end{IEEEeqnarray*} 
%\begin{subequations}
%\begin{align}
%    &find \:  \vec{P}\\
%\text{subject to}\nonumber \\
%    &{\gamma_n} \geq t, \: \: n\in \{ 1,\ldots, L \} \\
%    &\gamma_n \geq \Gamma_n,n \in \{1,\ldots,L \} \\
%    &\sum_{n=1}^{L}{P_n}\leq P_s.
%\end{align}
%\end{subequations} 

If we substitute (2) into (15), (15-b) and (15-c) can be written respectively as
\begin{equation}\label{20}
  P_nG_n-t\sum_{j=1}^{n-1}{P_j}G_n+tN_n\geq 0,
\end{equation}
and
\begin{equation}\label{21}
  P_nG_n-\Gamma_n \sum_{j=1}^{n-1}{P_j}G_n+\Gamma_n N_n\geq 0.
\end{equation} 
Further, we combine (16) and (17) through $\lambda_n=\max(t,\Gamma_n)$, which yields
\begin{equation}\label{22}
  P_nG_n-\lambda_n \sum_{j=1}^{n-1}{P_j}G_n+\lambda_n N_n\geq 0.
\end{equation}
Then, (15) can be reformulated as
\begin{IEEEeqnarray*}{clr}\label{eq:P}
\displaystyle {\text{find}}  & \quad \vec{P}_{L^ \star} \IEEEyesnumber \IEEEyessubnumber* \\
\text{s.t.}
   &\quad P_nG_n-\lambda_n \sum_{j=1}^{n-1}{P_j}G_n+\lambda_n N_n\geq 0, \\ \nonumber
   &\quad n\in \{ 1,\ldots, {L^ \star} \} \\
    &\quad \sum_{n=1}^{{L^ \star}}{P_n}\leq P_s.
\end{IEEEeqnarray*} 

%\begin{subequations}
%\begin{align}
%         & find \:  \vec{P}\\
%\text{subject to}\nonumber \\
%    &P_nG_n-\lambda_n \sum_{j=1}^{n-1}{P_j}G_n+\lambda_n N_n\geq 0 \\
%    &\sum_{i=1}^{L}{P_n}\leq P_s.
%\end{align}
%\end{subequations}

For any given value of $t$, $ \lambda_n$ has a specific value. Thus, (19) is a convex feasibility problem, since the inequality constraint functions are all linear. Let $\theta^{\star}$ denote the optimal value of the quasiconvex optimization problem (10). If (19) is feasible, i.e., there exists $\vec{P}_{L^\star}$ satisfying (19), we have $\theta^{\star} \geq t$. Otherwise, we can conclude $\theta^{\star} \leq t$. Therefore, we can check whether the optimal value $\theta^{\star}$ of a quasiconvex optimization problem is over or below a given value $t$ by solving the convex feasibility problem (19).

The observation above can be used as the basis of a simple algorithm for solving the quasiconvex optimization problem (19) via bisection, i.e., solving a convex feasibility problem at each step. Firstly, the problem is set to be feasible, e.g., we start with an interval $[l,u]$ known to contain the optimal value $\theta^{\star}$. Then, the convex feasibility problem is solved at its midpoint $t=(l+u)/2$, to determine whether the optimal value is in the lower or upper half of the interval, and update the interval accordingly. This produces a new interval, which also contains the optimal value, but has half the width of the initial interval. The progress is repeated until the width of the interval is small enough.\\

\begin{algorithm}
\caption{Optimal Method for Quasiconvex Optimization}\label{BM}
\begin{algorithmic}[1]
\State {\textbf{Initialize parameters.}}
\State \hspace{20pt} Given $l \leq \theta^{\star}, u \geq \theta^{\star}$, tolerance $\epsilon \geq 0$, where 
\State \hspace{20pt} $l=\min(\Gamma_n), u=\max(P_s/N_n), n \in \{1, \dots, {L^ \star}\}$;
\State {\textbf{repeat:}}
    \State \hspace{20pt} $t=(l+u)/2$;
    \State \hspace{20pt} Solve the convex problem (19);
    \State \hspace{20pt} If (19) is feasible, $u=t$; else $l=t$;
\State {\textbf{Until $u-l\leq \epsilon$}}
\end{algorithmic}
\end{algorithm}

The interval $[l,u]$ is guaranteed to contain $\theta^{\star}$, i.e., we have $l\leq \theta^{\star} \leq u$ at each step. It is obvious that $l=\min(\Gamma_n),n \in \{1, \dots, {L^ \star}\}$ can be used as the lower boundary, according to the SINR requirement. In terms of the upper boundary, the highest SINR the system can achieve should not exceed the value when all the power is allocated to a single SU. In each iteration, the interval is divided in two, i.e., bisected. Thus, the length of the interval after $k$ iterations is $2^{-k}(u-l)$, where $u-l$ is the length of the initial interval. It follows that exactly $\lceil log_2((u-l)/\epsilon)\rceil$iterations are required before the algorithm terminates. Each step involves solving the convex feasibility problem (19).

\subsubsection{An Analytical Solution Based on the Water-filling Scheme} The second problem can be solved using the optimal method for quasiconvex optimization. However, although the number of iterations is fixed for a given threshold, it is still computationally complex since each step requires solving the convex feasibility problem. In this section, we firstly prove the existence of a unique solution for this problem, and then propose a PA algorithm based on the water-filling scheme to obtain the optimal solution.

\paragraph{Existence of a unique solution}
We assume that the maximized minimum SINR among the admitted SUs via the full exploitation of the remaining power is $\theta^{\star}$. Accordingly, it is proven by contradiction in the following paragraph that $\gamma_n=\max(\theta^{\star},\Gamma_n),n = \{1, \ldots, {L^ \star}\}$, i.e., when $\Gamma_n \leq \theta^{\star}, \gamma_n=\theta^{\star}$; otherwise $\gamma_n=\Gamma_n$. Hence, once $\theta^{\star}$ is certain, the SINR for each admitted SU is updated accordingly from its targeted SINR. Then, based on (13), the power allocated to the SUs can be obtained, and their sum should satisfy $\sum_{n=1}^{{L^ \star}}{P_n}=P_s$. This equation only has one variable, i.e., $\theta^{\star}$. Furthermore, $P_s$ monotonically increases with $\theta^{\star}$. Since $P_s$ is fixed, $\theta^{\star}$ has a unique value.

\begin{IEEEproof}
Let us assume that for the $n$th SU, $\Gamma_n \leq \theta^{\star}, \gamma_n>\theta^{\star}$. Then, we can simply improve $\theta^{\star}$ by transferring some power from the $n$th SU to other SUs whose SINR equal to $\theta^{\star}$. This contradicts our premise that $\theta^{\star}$ is the maximized minimum SINR. Likewise, we can easily prove that for the $n$th SU with $\Gamma_n > \theta^{\star}$, $\gamma_n$ should equal $\Gamma_n$. 
\end{IEEEproof}

\paragraph{Proposed PA algorithm based on the water-filling scheme}
The above analysis shows the existence of a unique solution for the second problem. However, due to the operation of comparison between the targeted SINR and the maximized minimum SINR, i.e., $\max(\theta^{\star},\Gamma_n)$, the function between the overall power $P_s$ and $\theta^{\star}$ is piecewise. Indeed, this piecewise function can be calculated using the water-filling scheme. 

\iffalse
\begin{algorithm}
\caption{The Second PA Algorithm}\label{wf}
\begin{algorithmic}[1]
\State {\textbf{Initializing parameters.}}
\State \hspace{20pt} power vector: ${\bf{P}}=[P_1 \dots P_n \dots P_L]$; 
\State \hspace{20pt} remaining power: $P_r=P_s-\sum_{n=1}^{L}P_n$;
\State \hspace{20pt} SINR threshold vector: ${\bf{\Gamma}}=[\Gamma_1 \dots \Gamma_L]$; 
\State \hspace{20pt} maximized minimum SINR: $\theta$;
\State {\textbf{repeat:}}
    \State \hspace{20pt} $Pos=find(min({\bf{\Gamma}}))$; 
    \State \hspace{20pt} ${\bf{\Gamma'}}$: the elements of ${\bf{\Gamma}}$ except $min({\bf{\Gamma}})$; 
    \State \hspace{20pt} $\Gamma_{Pos}=min({\bf{\Gamma'}})$, and recalculate $P_{Pos}$ via (15);
    \State \hspace{20pt} $P_g=P_{Pos}-{\bf{P}}_{Pos}$; 
    \State \hspace{20pt} \bf{If} $P_g>P_r$ break;
    \State \hspace{20pt} \bf{Else} $P_r=P_r-P_g; {\bf{\Gamma}}={\bf{\Gamma'}}$;   
\State {\textbf{Until $u-l\leq \epsilon$}}
\end{algorithmic}
\end{algorithm}
\fi

\paragraph*{Procedure} After employing the initial PA algorithm, the SINR of each admitted SU is satisfied. There is some remaining power, which could be allocated to the admitted SUs. We first allocate the remaining power to the SU with lowest SINR. If the remaining power is large enough, the SINR of the SU with lowest SINR would reach the value of the one with the second lowest SINR. Then, power is assigned to the above two SUs so that their SINRs are equivalent to the SU with the third lowest SINR. The process repeats until there is no power left. Note that the bisection method may be required to obtain the value of $\theta^{\star}$, when it lies in the middle of two SINR thresholds of the admitted SUs.

\paragraph*{Optimality analysis} According to the PA process, we can conclude that the SINR of the SUs with extra power allocated should be equal. Moreover, the method is optimal. Assuming this is not optimal, then the optimal solution should provide  the SINR of the SUs with extra power allocated is not equal, i.e., at least the SINR of one SU is larger than another one. Since our objective is to maximize the minimum SINR of all admitted SUs, we can simply reallocate some power from the larger one to the smaller one to improve it. This conflicts with our hypothesis, and proves the optimality of the proposed PA scheme.

\begin{figure}
\centering
\includegraphics[width=0.5\textwidth]{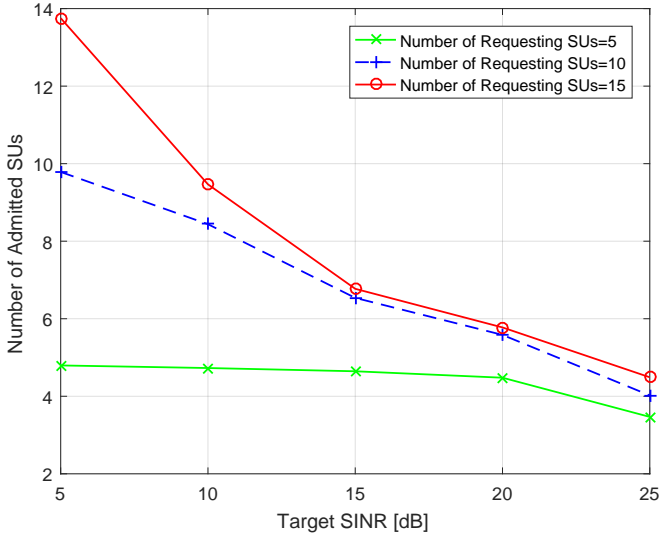}
\caption{Number of admitted SUs vs. targeted SINR.}
\end{figure}

\section{Simulation Results}
Simulations are run for a NOMA-based CRN shown in \text{Fig. 1}, with the cell radius of 500 m, and the BS located at the centre. The numerical results are obtained through averaging over $10^4$ simulation runs. During each simulation, SUs and PUs are randomly distributed in the area following uniform distribution. More exactly, their channel gains are modelled as $G_n=K\cdot 10^{\frac{H_n}{10}} D_n^{-4}$, $g_m=K\cdot 10^{\frac{h_m}{10}} d_m^{-4}$, where $D_n$ and $d_m$ are the corresponding distances, while $H_n$ and $h_m$ represent the lognormal shadowing, which are random Gaussian variables with zero mean and standard deviation equal to 6 dB. Additionally, system and transmission parameters e.g., antenna gain, carrier frequency, etc., are included in $K=10^3$. Moreover, we set $N_s=-120$ dBm, $I_m=-90$ dBm and $P_{max}=20$ dBm, where $N_s$ denotes the total noise and interference from all PUs for each SU.

\begin{figure}
\centering
\includegraphics[width=0.5\textwidth]{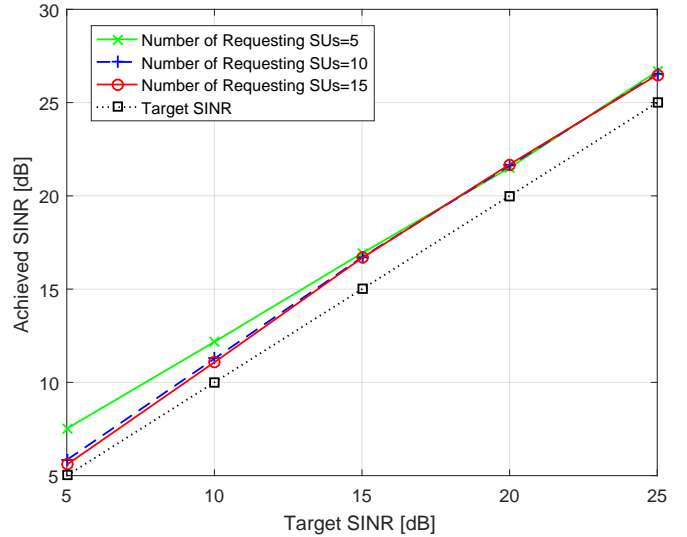}
\caption{Achieved SINR vs. targeted SINR.}
\end{figure}

The performance of the proposed two-phase PA scheme is investigated from two perspectives. First, the effectiveness of the initial PA process is studied. Fig.~2 shows the mean number of admitted SUs versus the targeted SINR, when the number of requesting SUs varies, i.e., $N=5, ~10$ or $15$. As the targeted SINR increases, the number of admitted SUs decreases for all three cases. However, the number of admitted SUs is quite high for all targeted SINRs. Particularly, when the targeted SINR is 5 dB, almost all requesting SUs are admitted for the three cases, which proves the effectiveness of the initial PA process. In addition, for any given targeted SINR, by comparing the three cases, one can observe that as the number of requesting SUs increases, the number of admitted SUs grows as well. This can be explained by the fact that as the number of requesting SUs increases, it is likely that more users will have better channel gains. According to (14), the increase in channel gains yields lower power consumption, and thus more users can be admitted. Even when the targeted SINR reaches 25 dB, about 4.5 users are admitted when $N=15$.

Following this, Fig. 3 compares the achieved SINR versus targeted SINR, when the targeted SINRs for all users are the same. For the case of $N=5$, about 1.5 dB increment is achieved for any given targeted SINR, which verifies the usefulness of the second PA process. As the number of requesting SUs grows, the growth in SINR declines in general. This is because the remaining power is divided by more admitted SUs. Particularly, when the targeted SINR is 5 and 10, the SINR increment for $N=10$ and $15$ is less than 1. This matches with Fig. 2, since for these two targeted SINR values, the admitted number of SUs are much larger than that of $N=5$. However, for the remaining three targeted SINR values, as the number of admitted SUs is almost the same for the three cases, a similar increase is obtained. In all, an average of 1 dB increment is obtained for the three cases, which validates the effectiveness of the proposed second PA process, when the targeted SINRs for all users are the same.

\begin{figure}
\centering
\includegraphics[width=0.5\textwidth]{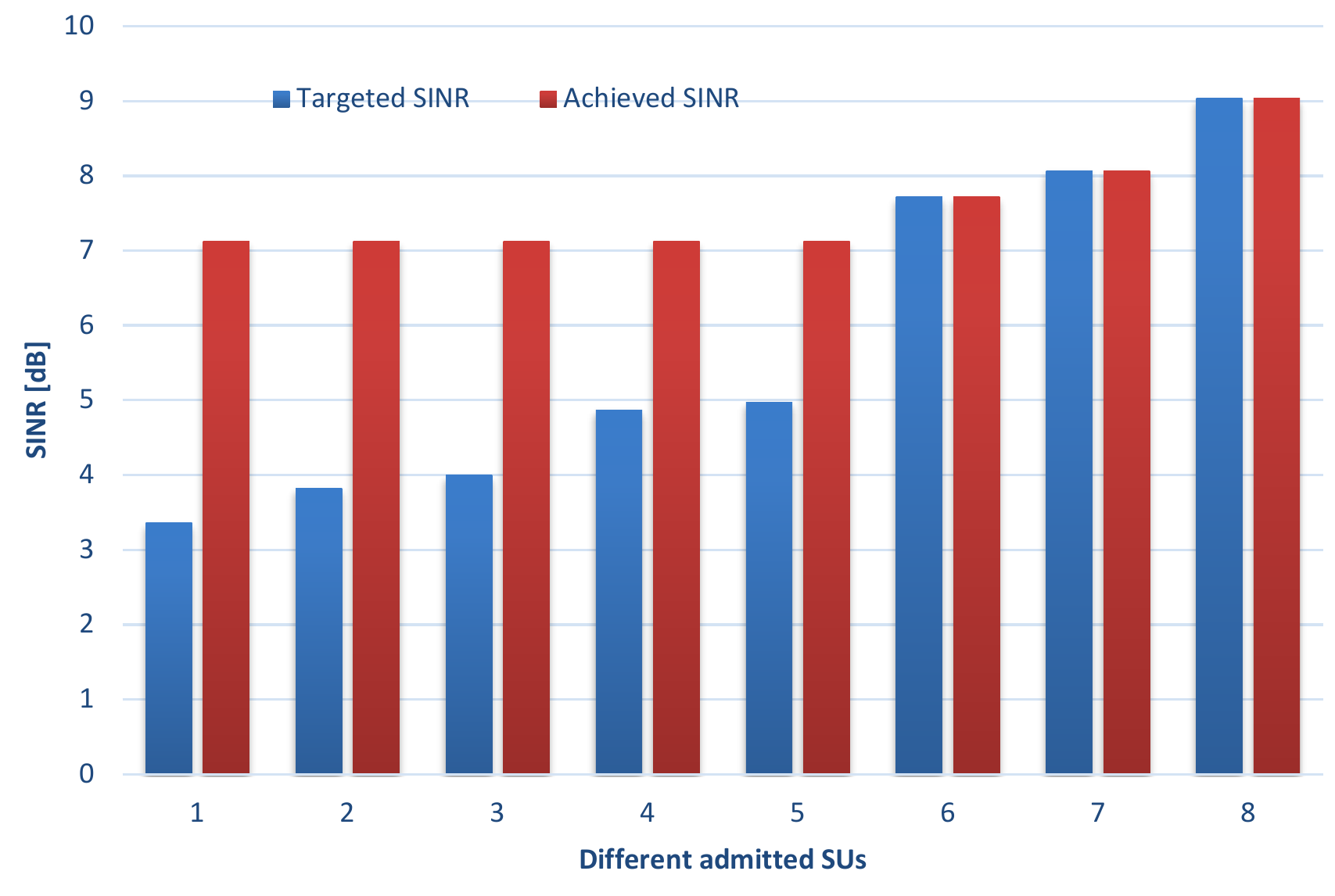}
\caption{Snapshot when the targeted SINR of each SU varies.}
\end{figure}

Lastly, we compare the achieved SINR versus the targeted SINR in Fig.~4, when the targeted SINR of each SU varies. Note that the requesting number of users is set to 15. Here, we cannot average the results, as the targeted SINR of each SU is different. Instead, a snapshot is given. Specifically, in Fig.~4, 8 SUs are admitted, among which the SINR values of 5 SUs are further increased, and for the other 3 SUs, they remain unchanged. On average, there is 1 dB increment. 

\section{CONCLUSIONS}
In this paper, PA for CRNs employing NOMA is considered. A two-phase PA scheme is proposed to maximize the number of admitted SUs and the throughput. Specifically, in the first phase, we apply the sequential and iterative PA algorithm to obtain the maximum number of admitted SUs. Following this, the second PA algorithm maximizes the minimum SINR among the admitted SUs. Simulation results show that the number of admitted SUs is large under different number of requesting SUs; and there is over 1 dB increment on average in the SINR of the admitted SUs, which verifies the effectiveness of the proposed two-phase PA scheme.

%\begin{figure}
%\centering
%\includegraphics[scale=0.5]{2.eps}
%\caption{An example of spectrum sharing among PUs and SUs.}
%\end{figure}
%
%\begin{figure}
%\centering
%\includegraphics[scale=0.5]{3.eps}
%\caption{An example of spectrum sharing among PUs and SUs.}
%\end{figure}
%
%\begin{figure}
%\centering
%\includegraphics[scale=0.5]{4.eps}
%\caption{An example of spectrum sharing among PUs and SUs.}
%\end{figure}

\section*{Acknowledgement}
This work has been supported in part by the Natural Sciences and Engineering Research Council of Canada (NSERC), through its Discovery program.

\bibliographystyle{IEEEtran}
\bibliography{IEEEabrv,bare-conf}

\begin{thebibliography}{10}
\providecommand{\url}[1]{#1}
\csname url@samestyle\endcsname
\providecommand{\newblock}{\relax}
\providecommand{\bibinfo}[2]{#2}
\providecommand{\BIBentrySTDinterwordspacing}{\spaceskip=0pt\relax}
\providecommand{\BIBentryALTinterwordstretchfactor}{4}
\providecommand{\BIBentryALTinterwordspacing}{\spaceskip=\fontdimen2\font plus
\BIBentryALTinterwordstretchfactor\fontdimen3\font minus
  \fontdimen4\font\relax}
\providecommand{\BIBforeignlanguage}[2]{{%
\expandafter\ifx\csname l@#1\endcsname\relax
\typeout{** WARNING: IEEEtran.bst: No hyphenation pattern has been}%
\typeout{** loaded for the language `#1'. Using the pattern for}%
\typeout{** the default language instead.}%
\else
\language=\csname l@#1\endcsname
\fi
#2}}
\providecommand{\BIBdecl}{\relax}
\BIBdecl

\bibitem{a1}
C.~V.~N. Index, ``Global mobile data traffic forecast update, 2012-2017,''
  \emph{Cisco white paper}, 2015.

\bibitem{a2}
N.~Alliance, ``5{G} white paper,'' \emph{Next Generation Mobile Networks, White
  paper}, 2015.

\bibitem{a3}
I.~F. Akyildiz, W.-Y. Lee, M.~C. Vuran, and S.~Mohanty, ``A survey on spectrum
  management in cognitive radio networks,'' \emph{IEEE Commun. Mag.}, vol.~46,
  no.~4, pp. 40--48, Apr. 2008.

\bibitem{a4}
G.~I. Tsiropoulos, O.~A. Dobre, M.~H. Ahmed, and K.~E. Baddour, ``Radio
  resource allocation techniques for efficient spectrum access in cognitive
  radio networks,'' \emph{IEEE Commun. Surv. Tuts.}, no.~99, pp. 82--95, Jan.
  2016.

\bibitem{a5}
G.~I. Tsiropoulos, Z.~Ming, O.~A. Dobre, and M.~H. Ahmed, ``A load-balancing
  semi-matching approach for resource allocation in cognitive radio networks,''
  in \emph{Proc. IEEE ICC}, 2016, pp. 1--6.

\bibitem{25}
G.~I. Tsiropoulos, A.~Yadav, M.~Zeng, and O.~A. Dobre, ``Cooperation in {5G}
  heterogeneous networks: Advanced spectrum access and d2d assisted
  communications,'' \emph{IEEE Wireless Commun. Mag.}, vol.~PP, no.~99, pp.
  1--1, Jun. 2017.

\bibitem{22}
S.~M.~R. Islam, M.~Zeng, and O.~A. Dobre, ``{NOMA} in 5{G} systems: Exciting
  possibilities for enhancing spectral efficiency,'' \emph{IEEE 5{G} Tech.
  Focus, \emph{vol. 1, no. 2, May 2017. [Online]. Available: 307
  http://5g.ieee.org/tech-focus}}.

\bibitem{23}
M.~Zeng, A.~Yadav, O.~A. Dobre, G.~I. Tsiropoulos, and H.~V. Poor, ``Capacity
  comparison between {MIMO-NOMA} and {MIMO-OMA} with multiple users in a
  cluster,'' \emph{IEEE J. Select. Areas Commun.}, vol.~PP, no.~99, pp. 1--1,
  Jun. 2017.

\bibitem{24}
------, ``On the sum rate of {MIMO-NOMA} and {MIMO-OMA} systems,'' \emph{IEEE
  Wireless Commun. Lett.}, vol.~PP, no.~99, pp. 1--1, Jun. 2017.

\bibitem{a19}
S.~M.~R. Islam, N.~Avazov, O.~A. Dobre, and K.~S. Kwak, ``Power-domain
  non-orthogonal multiple access {(NOMA)} in 5{G} systems: Potentials and
  challenges,'' \emph{IEEE Commun. Surv. Tuts.,}, vol.~19, no.~2, pp. 721--742,
  May 2017.

\bibitem{a10}
A.~Li, A.~Harada, and H.~Kayama, ``A novel low computational complexity power
  assignment method for non-orthogonal multiple access systems,'' \emph{IEICE
  Trans. Fundamentals}, vol. E97.A, no.~1, pp. 57--68, 2014.

\bibitem{a12}
N.~Otao, Y.~Kishiyama, and K.~Higuchi, ``Performance of non-orthogonal access
  with {SIC} in cellular downlink using proportional fair-based resource
  allocation,'' in \emph{Proc. IEEE ISWCS}, 2012, pp. 476--480.

\bibitem{a16}
Z.~Ding, R.~Schober, and H.~V. Poor, ``A general {MIMO} framework for {NOMA}
  downlink and uplink transmission based on signal alignment,'' \emph{IEEE
  Trans. Wireless Commun.}, vol.~15, no.~6, pp. 4438--4454, Jun. 2016.

\bibitem{a17}
Z.~Yang, Z.~Ding, P.~Fan, and N.~Al-Dhahir, ``A general power allocation scheme
  to guarantee quality of service in downlink and uplink {NOMA} systems,''
  \emph{IEEE Trans. Wireless Commun.}, vol.~15, no.~11, pp. 7244--7257, Nov.
  2016.

\bibitem{a21}
J.~Cui, Z.~Ding, and P.~Fan, ``A novel power allocation scheme under outage
  constraints in noma systems,'' \emph{IEEE Signal Process. Lett.}, vol.~23,
  no.~9, pp. 1226--1230, Sep. 2016.

\bibitem{a9}
S.~Timotheou and I.~Krikidis, ``Fairness for non-orthogonal multiple access in
  5{G} systems,'' \emph{IEEE Signal Process. Lett.}, vol.~22, no.~10, pp.
  1647--1651, Oct. 2015.

\bibitem{a14}
Z.~Ming, G.~I. Tsiropoulos, O.~A. Dobre, and M.~H. Ahmed, ``Power allocation
  for cognitive radio networks employing non-orthogonal multiple access,''
  \emph{in Proc. IEEE Globecom}, pp. 1--5, 2016.

\bibitem{a18}
L.~Dai, B.~Wang, Y.~Yuan, S.~Han, C.~l.~I, and Z.~Wang, ``Non-orthogonal
  multiple access for 5{G}: Solutions, challenges, opportunities, and future
  research trends,'' \emph{IEEE Commun. Mag.}, vol.~53, no.~9, pp. 74--81, Sep.
  2015.

\bibitem{a20}
Z.~Ding, Y.~Liu, J.~Choi, Q.~Sun, M.~Elkashlan, and H.~V. Poor, ``Application
  of non-orthogonal multiple access in {LTE} and 5{G} networks,'' \emph{IEEE
  Commun. Mag.}, vol.~55, no.~2, pp. 185--191, Feb. 2017.

\bibitem{a15}
S.~Boyd and L.~Vandenberghe, \emph{Convex Optimization}.\hskip 1em plus 0.5em
  minus 0.4em\relax Cambridge University Press, 2004.

\end{thebibliography}

%\bibliographystyle{IEEEtran}
%\begin{thebibliography}{1}


%\end{thebibliography}

\end{document}